# Reversible phase transformation and doubly-charged anions at the surface of simple cubic RbC$_{60}$

Roberto Macovez,[1*] Andrea Goldoni,[2] Luca Petaccia,[2]

Ingrid Marenne,[3] Paul A. Brühwiler,[4,5] and Petra Rudolf [1]

1 - Zernike Institute for Advanced Materials, University of Groningen, Nijenborgh 4, 9747 AG Groningen, the Netherlands

2 - Sincrotrone Trieste S.C.p.A., Strada Statale 14 Km 163.5, AREA Science Park, I-34012 Trieste, Italy

3 - L.I.S.E., Facultés Universitaires Notre-Dame de la Paix, Rue de Bruxelles 61, B-5000 Namur, Belgium

4 - Department of Physics, Uppsala University, Box 530, S-75121 Uppsala, Sweden

5 - Empa, Swiss Federal Laboratories for Materials Testing and Research, Lerchenfeldstrasse 5, CH-9014 St. Gallen, Switzerland

## ABSTRACT

The simple cubic phase of a RbC$_{60}$ thin film has been studied using photoelectron spectroscopy. The simple cubic-to-dimer transition is found to be reversible at the film surface. A sharp Fermi edge is observed and a lower limit of 0.5 eV is found for the surface Hubbard *U*, pointing to a strongly-correlated metallic character of thin-film simple cubic RbC$_{60}$. A molecular charge state is identified in the valence band and core level photoemission spectra which arises from C$_{60}^{2-}$ anions and contributes to the spectral intensity at the Fermi level.

---

[*]Corresponding author. Electronic address: roberto.macovez@icfo.es

Alkali fullerides, stoichiometric compounds of Buckminsterfullerene ($C_{60}$) with alkali atoms, exhibit a large variety of insulating, metallic, magnetic and superconducting behaviors as a function of alkali concentration [1-10]. Such a wide range of physical properties partially stems from the proximity of cubic alkali fullerides to a metal-insulator transition [1] governed by the interplay of intramolecular interactions, in particular the inter-electron repulsion [2] and the Jahn-Teller vibronic coupling [3,4]. While even-stoichiometry fullerides such as $Na_2C_{60}$ and $Rb_4C_{60}$ are diamagnetic Mott-Jahn-Teller insulators [5,6], odd-stoichiometry compounds are metals and in some cases even superconductors [7-10]. Understanding the unconventional conduction properties of these materials is crucial for the fundamental description of molecular solids and narrow-band correlated systems in general.

The lowest stoichiometry $AC_{60}$ fullerides (where $A$=K, Rb, Cs) display within a single composition a range of electronic properties and phases as a function of thermal treatment [11]. Two distinct cubic phases with differing metallic properties are observed: a face-centered cubic (fcc) structure of rapidly spinning molecules [12], which is thermodynamically stable above 400 K, and a metastable simple cubic (sc) phase, obtained by fast-cooling the fcc phase to below 100 K, which only differs from the high-temperature structure in that the orientation of the $C_{60}$ monomers is fixed with respect to the cubic axes [13]. Upon annealing to 200 K, the sc structure transforms irreversibly into metastable $(C_{60})_2^{2-}$ dimers [14,15], which can also be obtained by fast-cooling the fcc phase to below room temperature. A weakly-conducting phase of polymer chains is thermodynamically stable at and below room temperature [16,17].

Optical conductivity measurements on the fcc phase indicate metallic character [18], while NMR and ESR investigations showed a high degree of localization of the LUMO electrons on single molecules [12,17], suggesting that the bulk fcc phase is a strongly correlated metal. The bulk sc phase is a metal [13] with unusual features: in $CsC_{60}$ a partial spin gap was observed by NMR, and ascribed to the presence of a minority of stable doubly-charged molecules in spin-singlet state [19,20]. The same authors suggest that similar states might occur as short-lived charge fluctuations in the fcc phase [21]. Here, we study the sc phase of a $RbC_{60}$ thin film by means of photoelectron spectroscopy (PES). By means of this technique we have recently shown [22] that a surface charge reconstruction takes place in the fcc and dimer phases of this compound, which is here shown to be common also to the sc phase. In contrast to the bulk compound, the phase transformation from the sc phase to the dimer phase is shown to be reversible at the film surface, where the charge reconstruction takes place. We find evidence in the sc phase for the same charge states identified in NMR studies of bulk $CsC_{60}$, and show in particular that doubly-charged fullerenes contribute to the density of states at the Fermi level, which points to an active role of these charge states for the surface metallicity.

The PES experiments were performed at the SuperESCA beamline [23] of Elettra (base pressure of $5\times10^{-11}$ mbar) on a crystalline $RbC_{60}$ sample grown *in situ* on a Ag(100) single crystal via the standard distillation procedure [24], yielding a $C_{60}$-terminated fcc $RbC_{60}$(111) film [22]. The sc and dimer phases were obtained by fast-cooling from the fcc phase (see below). Core-level (valence band) spectra were acquired at 400eV (129eV) photon energy with 0.1eV (50meV) resolution. The binding energy scale was referenced to the Fermi level of the clean substrate. The photoelectrons were collected both at normal (0°) and grazing (70°) emission to establish possible differences between the surface and subsurface layers.

Figure 1(a) displays the temperature evolution of the frontier electronic states during the quench from the fcc to the dimer phase and as the temperature is further lowered. The feature closest to the Fermi level ($E_F$) in the spectrum of the fcc phase arises from the partial filling of the $C_{60}$

LUMO-derived band due to charge transfer from Rb. The transition to the insulating dimer phase is visible as the opening of a gap at $E_F$ in the density of states. The features around 1 eV in the spectra acquired at 230 K and 170 K arise from the highest filled molecular orbitals of the charged $(C_{60})_2^{2-}$ dimer [22,25,26]. These features could be observed in the temperature range between 250 K and 135 K.

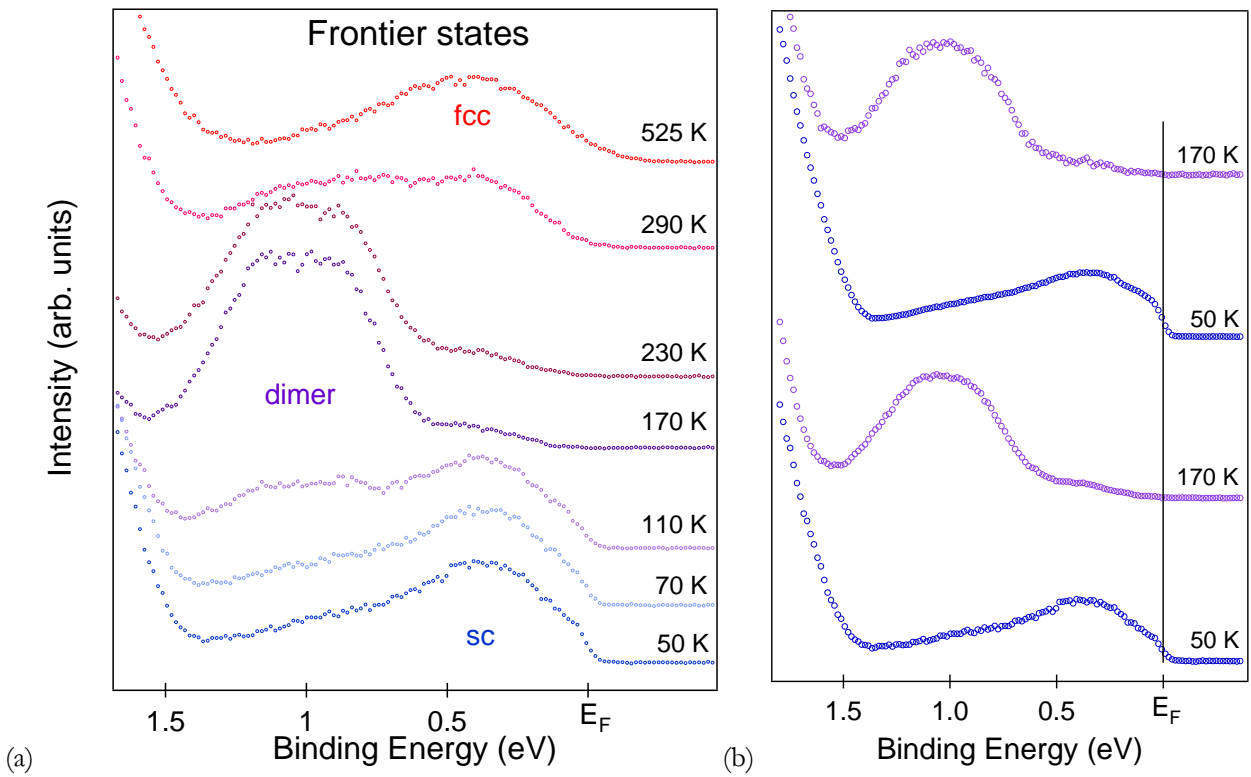

Figure 1. (a) Valence band photoemission spectra of RbC$_{60}$ evidencing the temperature-dependence of the states closest to $E_F$ during the quench from the fcc to the sc phase at a rate of 50 K per minute. (b) Spectra of the same binding energy region obtained while cycling the sample temperature between 170 and 50 K.

As the temperature is lowered below 135 K, the dimerized film undergoes a transition to a conducting phase characterized by a sharp Fermi edge. The sample could be annealed and cooled repeatedly through this transition, which shows its reversible character. This is shown in figure 1(b), which presents spectra taken while varying the sample temperature between 170 and 50 K twice, and was also verified in other synchrotron runs. The reversible transformation undergone by the dimer phase cannot be a transition to the polymer phase, since once formed, the polymer chains are stable up to temperatures high enough to recover the fcc phase. Moreover, the RbC$_{60}$ polymer phase undergoes a metal-to-insulator transition accompanied by magnetic ordering at 50 K [17,27], while our spectra (see lower spectrum in figure 1(a)) clearly point to a strongly metallic character of the low temperature phase, which would be consistent with the reported metallic character in the bulk [13]. The transition temperature from the dimer to the low-temperature phase observed in the spectra corresponds to the temperature at which the sc phase transforms (irreversibly) upon warming into the dimer phase in bulk RbC$_{60}$ (125 K) [13]. It should be noted that the inelastic mean free path of electrons in fullerides is of the order of the intermolecular distance or smaller [28], so that our PES data mainly reflect the character of the first and second molecular layer of the film [22]. Our observations thus establish the reversibility of the sc-dimer transition at the film surface, in contrast to the bulk case [14,15]. Remarkably, this entails that the covalent intermolecular bonds of the surface dimers are broken as the temperature is lowered.

The C 1*s* and valence band spectra of the two cubic phases are compared in fig. 2. The C 1*s* spectrum of the fcc phase consists of two components. In Fig. 2(a) a fit to this spectrum using two Doniach-Šunjić components separated by 0.65 eV is shown, with the addition of a Shirley background to account for inelastically-scattered electrons. The valence band spectrum of the same phase, apart from the LUMO-derived spectral feature, can similarly be reproduced as the sum of two C$_{60}$ lineshapes (Fig. 2(b)). We have previously assigned this two-component structure, which is common also to the dimer and polymer phases, to a charge reconstruction of the outermost molecular layer of the RbC$_{60}$ film, resulting in a surface layer consisting of 50% neutral and 50% charged molecules (see Ref. 22 and references therein).

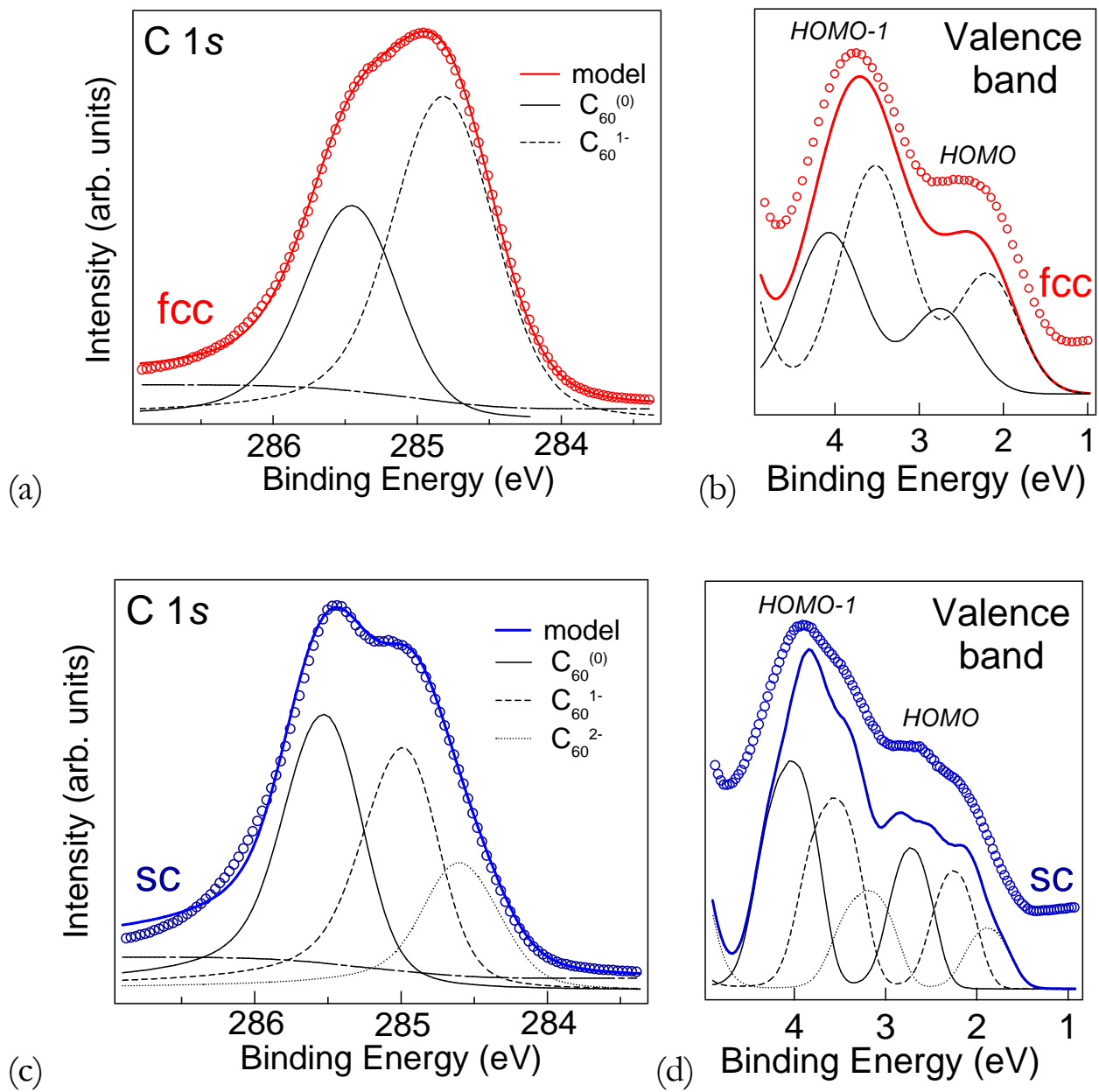

Figure 2. C 1$s$ photoemission spectra of the fcc (a) and sc (c) phases, and corresponding valence band photoemission spectra of the HOMO- and HOMO–1-derived region (b, d). A model in terms of different molecular charge states is shown for all spectra (see text).

The C 1s spectrum of the sc phase (Fig. 2(c)) is also characterized by a multiple-component structure. The presence of at least two components is clearly visible also in the corresponding valence band spectrum (Fig. 2(d)). In analogy with our study of the other phases [22] we associate the observation of two main components to the presence of both neutral and charged molecules. Simple considerations for fcc $RbC_{60}$ [22] and $K_3C_{60}$ [29,30] indicate that the component corresponding to the most negatively-charged molecules appears at lower binding energy due to inter-electron repulsion. It can be seen in panels (a) and (c) that the relative height of the two main C 1s features in the sc phase is reversed compared to the fcc phase, which would seem to indicate that the percentage of neutral fullerenes in the film has increased and that of $C_{60}^-$ ions decreased. This is accompanied by the emergence of a pronounced tail at low binding energy, which cannot be due to metallic screening of the core hole nor to inelastically-scattered electrons, since both effects only contribute an asymmetry towards higher binding energy. While it is possible to reproduce the core level and valence band spectra of the fcc phase with two components of the same width (Fig. 2(a, b)), a third, less intense, component must be introduced at low binding energy in the sc phase spectra (Fig. 2(c, d)). The relative intensity and energy position of the three components are consistent between the core and valence spectra, with an energy separation between adjacent components of approximately 0.5 eV, *i.e.* slightly smaller than that found in the fcc phase (0.65 eV).

The similar energy separation between the components and the shift of the minority component to lower binding energy lead us to conclude that the new component arises from doubly-charged fullerene molecules in the sc phase. The relative intensities, which reflect the abundance of each charge state near the surface, are consistent with expectations for a charge-reconstructed film in which approximately 15% of the molecules (both in the surface and subsurface layers) carry two electron charges. The formation of doubly-charged states must occur at the expense of singly-charged species, rationalizing the observed intensity redistribution in the spectra. These observations strongly support the proposal based on NMR data of a minority of doubly-charged molecules in the spin-singlet state in the sc phase of $CsC_{60}$ [19,20]. The parallelism between the two compounds may be expected given that bulk $RbC_{60}$ and $CsC_{60}$ have virtually identical phase diagrams [11]. The percentage of doubly-charged anions is the same in the two systems, consistent with the conjecture of Ref. 20 that the density of $C_{60}^{2-}$ anions is limited by their mutual Coulomb repulsion.

The energy separation between the components in the C 1$s$ spectrum represents a lower limit for the surface Hubbard $U$ in the cubic phases [22,29,30], hence establishing the strongly-correlated nature of the $RbC_{60}$ surface. As directly visible in the data of figure 2, the absolute energy position of each charge state varies slightly between the two phases, which presumably reflects a different contribution of the intermolecular potential [22] and electronic screening to the binding energy of each charge component. The dependence on the emission angle (not shown) of the C 1$s$ spectrum of the sc phase is similar to that of the fcc phase, with the relative intensity of the high binding energy component (neutral $C_{60}$ molecules) increasing at grazing emission, consistent with the expectation for a charge-reconstructed surface layer [22]. Due to the lower temperature, the width of the components is narrower in the sc spectrum than in that of the fcc phase. This appears more clearly in the Rb 3$d$ spectra of the same phases, shown in figure 3(a). In both cubic phases only one spin-orbit doublet is observed, consistent with the expectations for cubic $AC_{60}$ fullerides [22]. In the sc spectrum the asymmetry of the features is more pronounced, in agreement with the more metallic character of this phase.

Figure 3(b) shows detailed spectra of the conduction band(s) in the fcc and sc phases of $RbC_{60}$ acquired at the same times as the spectra of the corresponding phases presented in figures 2 and 3a. While the sc spectrum displays a sharp Fermi edge followed by a maximum at 0.4 eV, the fcc spectrum consists of a broad feature centered at roughly the same binding energy [22,25,31] with a tail extending well above $E_F$ and decaying more slowly than the Fermi-Dirac distribution at the measuring temperature of 525 K [22]. This difference in line shape is reminiscent of the temperature dependence of the LUMO-derived states in $K_3C_{60}$ and $Rb_3C_{60}$ [32,33], where a smooth evolution is observed between a sharp, structured Fermi edge at low temperatures and a broad featureless LUMO profile at high temperatures, accompanied by a corresponding broadening of the other valence band features and the appearance of a tail above $E_F$ [32,33]. This congruence indicates that the non-Fermi liquid profile at high temperatures in odd stoichiometries is due to the strong electron-phonon coupling, and that the temperature dependence of the Fermi edge is mainly a consequence of the increased spectral weight of phonon satellites at higher temperatures.

The sc spectrum did not show any dependence on the emission angle of photoelectrons (see fig. 3b). The observation of a clear Fermi edge in both emission geometries demonstrates the metallic nature of the sc phase also in thin-film form and indicates that the metallic profile is common to the surface and bulk of the sample. The stronger intensity near 1 eV in the sc spectrum might be due to a more important contribution of inelastically-scattered electrons or spectral satellites, and partially also to the remnant presence of dimers.

Only charged species contribute to the LUMO-derived intensity, in contrast with the other electronic levels. From the relative weight of the $C_{60}^{2-}$ and $C_{60}^{1-}$ components in the C 1$s$ spectrum, we estimate that the doubly-charged states contribute approximately 40% of the total

LUMO intensity, with the majority contribution (60%) coming from 1– states. The energy position of these two components should follow that observed in the C 1$s$ and valence levels of the same phase, as it mainly reflects the contribution of the on-site Coulomb repulsion. Since the LUMO spectral maximum occurs roughly at the same energy as in the fcc phase, this prominent feature may be assigned to $C_{60}^-$ species, leading to the remarkable observation that the spectral weight at $E_F$ stems from doubly-charged molecules, in line with the analysis of the LUMO-derived spectrum of $K_3C_{60}$ in Ref. 29.

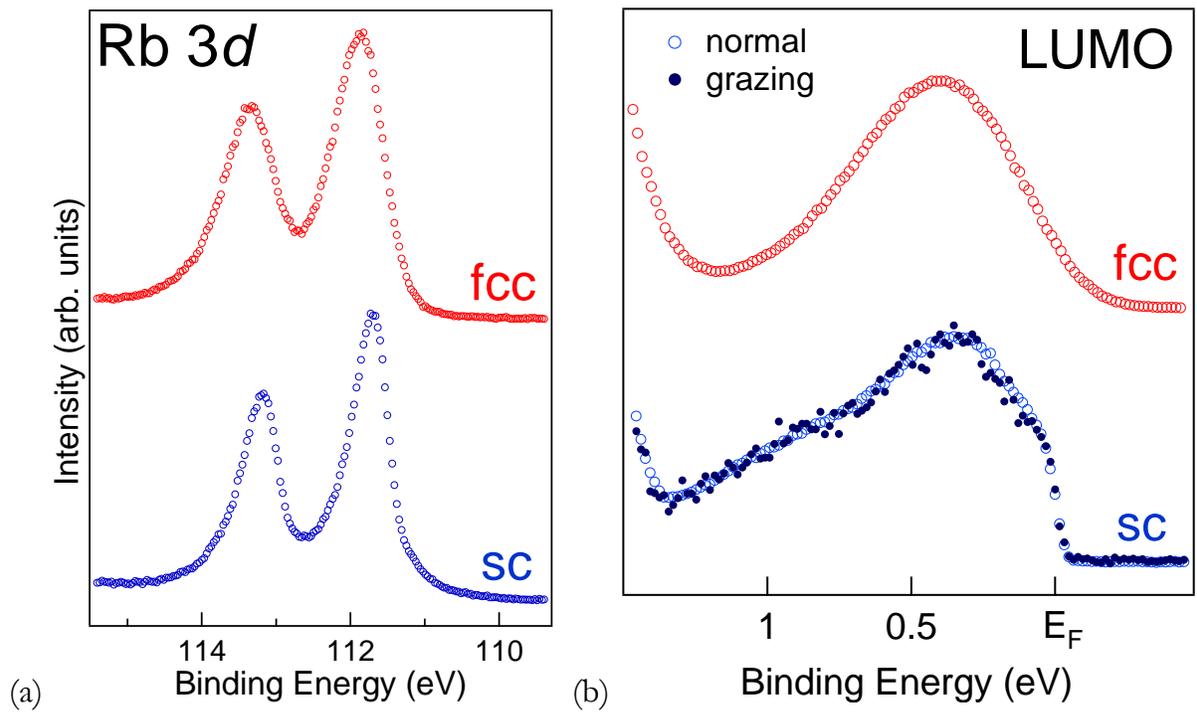

Figure 3. Comparison between the Rb 3$d$ (a) and LUMO-derived band (b) photoemission spectra of the fcc and sc phases. Open circles correspond to data acquired at normal emission, while dots at grazing emission.

Given that the value of the surface Hubbard $U$ found here (>0.5 eV) is much larger than the calculated value of the Jahn-Teller pairing energy favoring the formation of $C_{60}^{2-}$ anions [3,4,20], it may be surprising that the 2– charge state are observed in such a correlated system.

Nonetheless, a series of NMR studies has pointed to the occurrence of non-stoichiometric charge states in many fullerene compounds with a lifetime ranging between few and several tens of fs (depending on the phase, temperature and applied pressure) [6,21,34], except in bulk sc $CsC_{60}$ where the lifetime is even of the order of seconds. The stabilization of the 2– anions in the sc phase of $CsC_{60}$ was tentatively justified postulating the existence, besides the Jahn-Teller contribution, of a trapping potential associated with the orientational disorder typical of the sc phase, based on the observed correlation between the spin singlet fraction and the percentage of molecules with a minority orientation with respect to the cubic axes [20]. Distinct molecular orientations were observed also at the surface of pristine fullerite [35], where rotational phase transitions analogous to those of bulk $C_{60}$ have been identified [36,37]. Interestingly, the percentage of molecules in the minority orientation at the $C_{60}(111)$ surface (one out of four, or 25%) is close to the population of doubly-charged anions found in the present study, which similarly suggest that structural degrees of freedom could play an important role.

The fact that the doubly-charged states contribute the spectral intensity at $E_F$ implies that they are actively involved in the surface metallicity. This contrasts with the bulk case where long localization times are found [20], and suggests a shorter fluctuation time – possibly close to the values observed in other fullerenes [6,21,34], which would still allow for their observation on the PES time (tens of attoseconds). A shorter localization time at the surface might arise due to the poorer surface screening [38] hindering the stabilization of 2– states on long times, or to the lower energy barrier to rotational motion (especially for minority orientation molecules [36,37]) reducing the hopping time [20]. The simultaneous observation in the $RbC_{60}$ spectra of several molecular charges and the contribution of the doubly-charged state to the spectral intensity at $E_F$ prove the exquisitely molecular character of this correlated metallic fulleride, and provide a strong indication that the surface metallicity is mediated by molecular charge fluctuations. A quantitative understanding of this phenomenon would help shed light on the unconventional nature of conduction and superconductivity in $C_{60}$ solids. The peculiar features of the $RbC_{60}$ surface, in particular the occurrence of fluctuations in the charge-reconstructed termination layer, and the reversibility of the sc-to-dimer phase transformation, are emblematic of the subtle interplay of microscopic interactions which underlie the electronic and structural properties of fullerides.

ACKNOWLEDGEMENTS

The authors wish to thank Thomas Pichler and Lisbeth Kjeldgaard for help with preliminary experiments. The measurements at ELETTRA were supported by the "Access to Research Infrastructure" action of the improving Human Potential Program (ARI) of the EU. Additional support came from the Dutch Foundation for Fundamental Research on Matter (FOM), from the Breedtestrategie program of the University of Groningen, and the Caramel Consortium, the latter in turn being supported by Stiftelsen för Strategisk Forskning, as well as Vetenskapsrådet and Göran Gustafsons Stiftelse. I.M. acknowledges for financial support the FRIA (Belgium) and A,G. the MIUR (Italy) support by PRIN-2006022847.